\documentclass{article}

\usepackage{arxiv}
\usepackage[utf8]{inputenc} 
\usepackage[T1]{fontenc}    
\DeclareUnicodeCharacter{0301}{\'}
\usepackage{hyperref}       
\usepackage{url}            
\usepackage{booktabs}       
\usepackage{amsfonts}       
\usepackage{nicefrac}       
\usepackage{microtype}      
\usepackage{lipsum}
\usepackage{graphicx}
\graphicspath{ {./images/} }

\usepackage{amsmath}
\usepackage{amsfonts}
\usepackage{amssymb}

\usepackage{caption}
\title{Physics-Informed Neural Networks as Fast Surrogate Models for Electrochemical Flow Reactors}

\author{
 Eric Fernández-García \\
  School of Engineering and Sciences\\
  Tecnológico de Monterrey\\
  Monterrey, Mexico\\
  \texttt{eric.fernandez@tec.mx} \\
   \And
 Miguel Modestino\\
 Department of Chemical and Biomolecular Engineering\\
 New York University\\
  Brooklyn, USA\\
  \texttt{modestino@nyu.edu} \\
  \And
 Sergio Maldonado\\
  Faculty of Engineering and Physical Sciences\\
  University of Southampton\\
  Southampton, UK\\
  \texttt{s.maldonado@soton.ac.uk} \\
}

\begin{document}
\maketitle
\begin{abstract}
This work presents a physics-informed neural network (PINN) for modeling a transient two-dimensional electrochemical flow reactor with diffusion, migration, convection, and nonlinear anodic Butler--Volmer kinetics. The model is trained without labeled concentration data by embedding the governing transport equation and all initial and boundary conditions into a composite loss function. Spatial and temporal coordinates together with anodic overpotential, temperature, inlet concentration, maximum flow velocity, and diffusivity are used as inputs, allowing the network to predict concentration fields over a broad operating domain. Validation against finite-difference-based solutions shows strong agreement for representative transient and near-steady cases, with a mean relative space--time error of $(9.99 \pm 0.65)\times10^{-3}$ (sub-percent level) across the conditioning domain. PINN inference is faster than a traditional finite difference solver by a factor of $5.34$, thus reducing runtime by $81.3\%$. Generalization tests further show that the surrogate remains robust under in-domain boundary-focused sampling and controlled extrapolation, although anodic overpotential is the most challenging parameter due to its exponential effect on interfacial kinetics. The results indicate that physics-informed neural networks can serve as accurate and efficient parametric surrogates for electrochemical transport problems and provide a foundation for low-computational-cost digital-twin modeling of electrochemical flow reactors.
\end{abstract}


\section{Introduction}
Electrochemical flow reactors play a central role in modern electrosynthesis, energy conversion, and environmental technologies, where the spatial distribution of ionic species directly governs reaction rates, current efficiency, and product selectivity \cite{noel2019fundamentals}. Accurate prediction of concentration fields and electrochemical performance is therefore essential for optimizing reactor design, minimizing overpotentials, and enhancing the utilization of electroactive species. As highlighted in recent analyses of flow-based electrochemical systems, transport phenomena often dictate performance because diffusion, migration, and the geometry of narrow electrode gaps strongly influence how species reach reactive surfaces, and their coupling with electric fields, hydrodynamics, and electrode kinetics leads to complex physicochemical behavior \cite{noel2019fundamentals}.

Traditional numerical simulations, including finite-difference, finite-volume, and finite-element methods, have been widely applied to model diffusion, migration, and convection in electrochemical systems. These techniques have been validated across diverse geometries, such as parallel-plate reactors and backward-facing step configurations in multi-ion transport studies \cite{georgiadou1997finite}, channel flow cells used to investigate coupled homogeneous and heterogeneous kinetics \cite{stevens2000computational}, microreactors operating under laminar and interfacial transport regimes \cite{plazl2010modeling}, and even turbulent parallel-plate flows where multi-ion transport is strongly affected by high Schmidt numbers \cite{nelissen2004multi}. Although these numerical approaches achieve high accuracy, they often require fine spatial discretization, complex meshing strategies, and substantial computational resources, particularly when exploring wide ranges of operating conditions or solving fully coupled electrochemical systems. This computational burden limits their practicality when rapid evaluation of experimental scenarios or reactor configurations is needed. Furthermore, modeling realistic 3-dimensional reactor architectures and operation under dynamic conditions can often become computationally intractable, limiting the possibility to optimize electrochemical reactors in silico. 

At the same time, the electrochemical engineering community faces increasing demand for tools capable of screening operating conditions, understanding transport–reaction interactions, and guiding experimental design to achieve high-performing reactors that can advance chemical manufacturing practices. Flow reactors operating under laminar or microfluidic regimes are particularly attractive due to their well-defined hydrodynamics, enhanced mass transfer, and scalability through parallelization \cite{modestino2016potential}. However, even in these controlled geometries, predicting concentration gradients under varying flow velocities, electrode kinetics, and electric fields remains computationally intensive. Recent analyses of flow-based electrochemical systems have shown that subtle modifications in channel geometry, hydrodynamic structure, or electrode configuration can strongly influence species transport, current density, and overall reactor performance, as demonstrated in studies of gas diffusion electrodes and microfluidic electrochemical architectures \cite{souza2024fluid}. Modeling porous electrodes, such as those employed in industrial electrosynthesis reactors, is even more computationally demanding, as the resolution of multiscale transport, distributed reaction sites, and complex pore-scale hydrodynamics requires fine spatial discretization and tightly coupled solvers, often pushing conventional simulations to the limits of tractability \cite{bui2022continuum}.

Given these challenges, there is a growing interest in alternatives to conventional numerical solvers that can provide fast, flexible, and generalizable predictions. Physics-informed neural networks (PINNs), originally introduced by Raissi et al. to solve classical benchmark equations such as Burger, Schrödinger, and Allen–Cahn \cite{raissi2019physics}, have emerged as a powerful framework for solving partial differential equations without spatial discretization by embedding the governing laws directly into the loss function. In electrochemical systems, initial demonstrations include simulations of voltammetry, reconstruction of diffusion fields, and forward–inverse analysis of coupled kinetics \cite{zhang2025multitask,chen2022predicting}. These studies show that PINNs can efficiently handle high-dimensional domains and nonlinear boundary conditions while maintaining physical consistency.

Building on these developments, this work introduces a generalized PINN framework that is capable of predicting steady and transient concentration fields across a broad range of electrochemical operating conditions, using a simplified yet representative electrochemical flow reactor geometry as a model system. The proposed network incorporates the governing equations of mass conservation with diffusion, migration, and convection, as well as Butler–Volmer electrode kinetics, enabling predictive modeling without requiring labeled data. By integrating key physicochemical parameters as inputs, the model allows rapid exploration of experimental conditions that would otherwise require repeated numerical simulations.

The main objective of this study is to explore the potential of PINNs to fundamentally accelerate the modeling of electrochemical flow reactors. Once trained, the proposed PINN serves as a fast surrogate that can function as a digital twin of the reactor, enabling rapid exploration of geometries, operating conditions, and physicochemical parameters entirely in silico. This capability opens the door to computationally driven optimization of reactor design and operation without relying on prolonged and costly experimental campaigns, shifting the paradigm from iterative laboratory trial-and-error toward predictive, simulation-guided development of electrochemical systems.

\section{Problem formulation}

\begin{figure}[h!]
    \centering
    \includegraphics[width=0.7\linewidth]{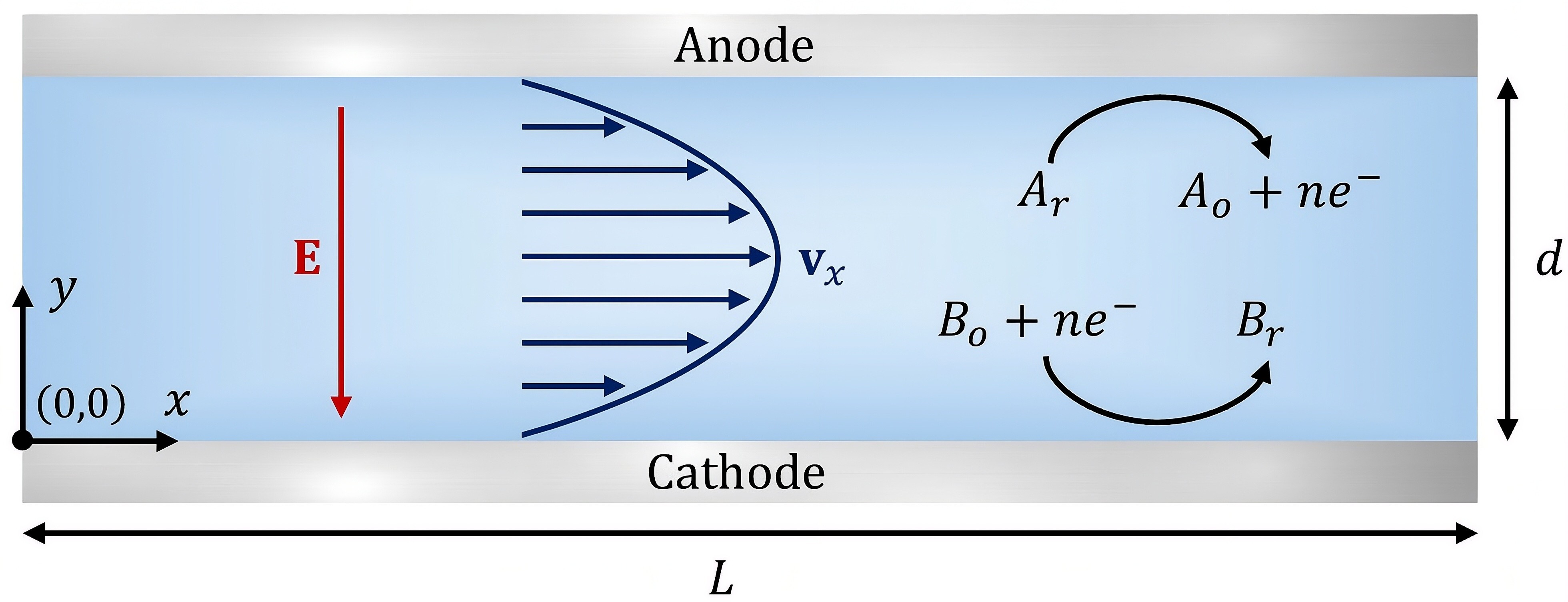}
\caption{Geometry and physical configuration of the two-dimensional electrochemical flow reactor. 
The rectangular channel ($L \times d$) is bounded by a planar anode at $y=d$ and a cathode at $y=0$. 
An externally imposed potential difference generates a spatially uniform electric field $\mathbf{E} = -\nabla \Phi$ directed downward. 
A pressure-driven laminar Poiseuille flow develops along the $x$-direction with characteristic velocity scale $v_0$, yielding a parabolic profile with maximum velocity $v_{\max} = v_0/4$ at the channel centerline. 
Oxidation of species $A_r$ to $A_o$ occurs at the anode, whereas reduction of species $B_o$ to $B_r$ takes place at the cathode.}
    \label{fig:S1}
\end{figure}

The system under consideration is a two-dimensional electrochemical flow reactor of length $L$ and channel height $d$, in which a fully developed laminar profile governs the convective transport between two parallel electrodes. The streamwise velocity, $\mathbf{v}_x$, follows the classical Poiseuille solution
\begin{equation}
\mathbf{v}_x\!\left(y\right) = v_0 \frac{y}{d}\left(1 - \frac{y}{d}\right)\hat{\mathbf{x}},
\label{eq:poiseuille}
\end{equation}
representing incompressible flow in a narrow channel, where $v_0$ is a characteristic velocity scale such that the maximum velocity is $v_{\max} = v_0/4$, attained at the channel centerline $y = d/2$, and $y \in [0,d]$ is the transverse coordinate measured from the cathode \cite{white2011fluid}. Species $A$ and $B$ undergo oxidation and reduction at the anode ($y=d$) and cathode ($y=0$), respectively, while both are advected along the $x$-direction from the inlet at $x=0$, as illustrated in Figure~\ref{fig:S1}.

Transport in dilute electrolytes is governed by the combined effects of migration, diffusion, and convection, described by the Nernst--Planck flux:
\begin{equation}
\mathbf{N}_i 
= -z_i u_i F C_i \nabla \Phi 
- D_i \nabla C_i 
+ C_i \mathbf{v}_x,
\label{eq:nernst_planck}
\end{equation}
where $C_i$ is the molar concentration of species $i$, $z_i$ its charge number, $u_i$ the ionic mobility, $F$ the Faraday constant, $D_i$ the diffusion coefficient, $\Phi$ the electric potential, and $\mathbf{v}_x$ the prescribed laminar velocity field. In the present configuration, the index $i$ corresponds to $i \in \left\{A, B\right\}$.

The temporal evolution of each species is governed by the local mass conservation equation,
\begin{equation}
\frac{\partial C_i}{\partial t} = - \nabla \cdot \mathbf{N}_i + R_i,
\label{eq:mass_conservation}
\end{equation}
where $R_i$ represents a volumetric source term associated with homogeneous reactions. In the present formulation, bulk reactions are neglected, such that $R_i = 0$.

Migration originates from the electrostatic term $-z_i F \nabla \Phi$, diffusion is driven by concentration gradients, and convection arises from the bulk fluid velocity field \cite{fuller2018electrochemical}. In this configuration, the electric field is assumed spatially uniform and oriented along the transverse direction,
\begin{equation}
\mathbf{E}=-\nabla \Phi.
\end{equation}
This approximation corresponds to the electroneutral bulk limit with uniform ionic conductivity $\sigma$, for which the electric potential satisfies
\begin{equation}
\nabla \cdot\left(\sigma \nabla \Phi\right)=0.
\end{equation}
For constant conductivity this reduces to Laplace's equation,
\begin{equation}
\nabla^2 \Phi=0.
\label{eq:laplace}
\end{equation}
In a parallel-plate configuration where potential variations are primarily transverse, i.e. $\Phi=\Phi\left(y\right)$, Eq.~\eqref{eq:laplace} reduces to a one-dimensional form, yielding a linear potential profile and therefore a spatially uniform electric field across the channel. Such a constant-field approximation is commonly adopted in electrochemical transport modeling when migration is driven primarily by an externally imposed potential gradient rather than by local charge accumulation \cite{bard2022electrochemical,newman2021electrochemical}.

Under these conditions, the electrostatic contribution to the Nernst--Planck flux can be interpreted as an effective electromigration velocity. Accordingly, for each ionic species $i$, the migration velocity is defined as
\begin{equation}
\mathbf{v}_{m,i}\equiv z_i u_i F \nabla \Phi,
\label{eq:migration_velocity}
\end{equation}
such that the migrative contribution to the molar flux can be expressed compactly as
$\mathbf{N}_{i}^{\mathrm{mig}} = - C_i \mathbf{v}_{m,i}$.

Substitution of the Nernst--Planck flux expression (Eq.~\ref{eq:nernst_planck}) into the mass conservation equation (Eq.~\ref{eq:mass_conservation}), together with the definition of the migration velocity in Eq.~\eqref{eq:migration_velocity} and the constant-field assumption, yields the governing transport equation for each species,
\begin{equation}
\frac{\partial C_i}{\partial t}
=
\nabla C_i \cdot \left(
\mathbf{v}_{m,i}
-
\mathbf{v}_x
\right)
+
D_i \nabla^2 C_i,
\label{eq:species_balance}
\end{equation}
where the advective and electromigrative contributions appear as first-order transport terms, while diffusion is represented by the Laplacian operator. Under the assumptions of electroneutrality, negligible homogeneous reactions in the bulk electrolyte, and a spatially uniform electric potential gradient, the transport equations become formally decoupled within the domain \cite{fuller2018electrochemical}. In this framework, coupling between electroactive species arises only through interfacial electrochemical reactions at the electrode boundaries.

The anodic and cathodic reactions are described by Butler--Volmer-type expressions, and the interfacial current density is formally written as
\begin{equation}
j = j_a = - j_c,
\label{eq:current_convention}
\end{equation}
\begin{equation}
j_a =
j_{a,0}
\left(\frac{C_{A,s}}{C_{A,0}}\right)^{\gamma_a}
\exp\!\left(\frac{\alpha_a F \eta_a}{RT}\right),
\label{eq:BV_anode}
\end{equation}
\begin{equation}
j_c =
j_{c,0}
\left(\frac{C_{B,s}}{C_{B,0}}\right)^{\gamma_c}
\exp\!\left(\frac{\alpha_c F \eta_c}{RT}\right),
\label{eq:BV_cathode}
\end{equation}
where the subscripts $a$ and $c$ refer to the anode and cathode, respectively. Here, $j_{a,0}$ and $j_{c,0}$ denote the reference exchange current densities, $C_{i,s}$ the interfacial concentration of the electroactive species, and $C_{i,0}$ the corresponding bulk reference concentration. Furthermore, $\gamma_a$ and $\gamma_c$ are the reaction orders, $\alpha_a$ and $\alpha_c$ the charge-transfer coefficients, and $\eta_a$ and $\eta_c$ the electrode overpotentials. $R$ is the universal gas constant and $T$ the absolute temperature. The exponential dependence on $\eta$ reflects the modulation of activation barriers by the applied electrode potential, consistent with classical charge-transfer theory.

The interfacial flux boundary conditions at the electrodes are defined by the anodic $\left(j_a\right)$ and cathodic $\left(j_c\right)$ current densities and the stoichiometry of the reactions as follows,
\begin{align}
y = 0:\quad
&\begin{cases}
\mathbf{N}_A \cdot \hat{\mathbf{n}} = 0, \\
\mathbf{N}_B \cdot \hat{\mathbf{n}} = - j_c/\left(nF\right),
\end{cases}
\label{eq:bc_cathode}
\\[4pt]
y = d:\quad
&\begin{cases}
\mathbf{N}_A \cdot \hat{\mathbf{n}} = j_a/\left(nF\right), \\
\mathbf{N}_B \cdot \hat{\mathbf{n}} = 0,
\end{cases}
\label{eq:bc_anode}
\end{align}
where $\hat{\mathbf{n}}$ denotes the outward unit vector normal to the electrode surface and
$n$ is the number of electrons transferred per electrochemical reaction.

At the inlet, the species concentrations are fixed at their bulk reference values,
whereas a zero-gradient condition is imposed at the outlet:
\begin{align}
x = 0: &\quad C_i = C_{i,0},
\label{eq:bc_inlet}
\\
x = L: &\quad \frac{\partial C_i}{\partial x} = 0.
\label{eq:bc_outlet}
\end{align}
The initial condition assumes a spatially uniform state,
\begin{equation}
t = 0: \quad C_i = C_{i,0}.
\label{eq:ic}
\end{equation}

Although the full electrochemical system formally involves two reactive species, the present model adopts a one-species transport formulation in which only the concentration field of species $A$ is explicitly solved. This reduction is justified under conditions where the counter electrode reaction is non-limiting, either because the complementary species is present in large excess or because its heterogeneous charge-transfer kinetics are sufficiently fast. Under such circumstances, the overall current is controlled exclusively by the transport and interfacial kinetics of species $A$, a common simplification in electrochemical modeling when one half-reaction dominates the response \cite{bard2022electrochemical,newman2021electrochemical}. Accordingly, the interfacial flux is determined solely from the anodic Butler-Volmer contribution, while the cathode is treated as impermeable to species $A$. The formal identity $j=j_a=-j_c$ is retained for thermodynamic consistency, although only $j_a$ enters the reduced transport description.

This reduced description preserves the essential coupling between transport and anodic reaction kinetics while significantly simplifying the computational model. The resulting system can be solved using conventional numerical approaches, such as finite-difference methods, which we employ here to produce the reference solutions to appraise the proposed PINN. The extension to multicomponent systems follows directly from the same transport formalism, as each species satisfies an equation of the form of Eq.~(\ref{eq:species_balance}) with coupling arising through charge conservation and interfacial kinetics. A full multicomponent implementation, while formally straightforward, introduces additional numerical stiffness and is therefore deferred to future work.

\section{Methods} \label{PINN}
\subsection{PINN formulation}

The PINN developed in this work is designed to approximate the solution of the coupled transport--reaction problem for species $A$ under experimentally relevant operating conditions. The network input vector $\mathbf{x}$ consists of the spatial and temporal coordinates together with a set of controllable physicochemical and operating parameters, namely:
\begin{equation}
\mathbf{x} = \left(x, y, t, \boldsymbol{\theta}\right),
\qquad
\boldsymbol{\theta} = \left(\eta_a, T, C_{A,0}, v_0, D_A\right),
\end{equation}
where $\boldsymbol{\theta}$ denotes the vector of conditioning parameters that define the operating state of the system. For convenience, we also introduce the set
\begin{equation}
\Theta = \{\eta_a, T, C_{A,0}, v_0, D_A\},
\end{equation}
so that individual parameters can be written compactly as $\theta \in \Theta$.

The neural network, denoted by $\aleph$, provides a continuous approximation of the concentration field of species $A$ over the spatio--temporal domain according to
\begin{equation}
C_A\!\left(\mathbf{x}\right) \approx \widetilde{C}_A\!\left(\mathbf{x}\right) = \aleph\!\left(\mathbf{x}; \boldsymbol{\Gamma}\right),
\end{equation}
where $\widetilde{C}_A\left(\mathbf{x}\right)$ represents the PINN prediction and $\boldsymbol{\Gamma}$ collectively denotes all trainable parameters of the network, including weights and biases.

The network trainable parameters are obtained by minimizing a composite loss function that enforces the governing transport equation, the initial condition, and the boundary conditions at the inlet, outlet, anode, and cathode. The total loss function is defined as
\begin{equation}
\mathfrak{L}
=
\lambda_1 \mathfrak{L}_{\Omega}
+
\lambda_2 \mathfrak{L}_0
+
\lambda_3 \mathfrak{L}_{\mathrm{D}}
+
\lambda_4 \mathfrak{L}_{\mathrm{B}}
+
\lambda_5 \mathfrak{L}_{a}
+
\lambda_6 \mathfrak{L}_{c},
\label{eq:objective_function}
\end{equation}
where $\mathfrak{L}$ is a scalar objective function and
$\boldsymbol{\lambda} = \left(\lambda_1,\ldots,\lambda_6\right)$ denotes the set of weighting coefficients controlling the relative contribution of each loss term \cite{wang2021understanding}. 
Specifically, $\mathfrak{L}_{\Omega}$ penalizes the residual of the species mass-conservation equation over the spatio--temporal domain, whereas $\mathfrak{L}_{0}$ enforces the initial condition at $t=0$. The loss terms $\mathfrak{L}_{\mathrm{D}}$ and $\mathfrak{L}_{\mathrm{B}}$ correspond to the inlet and outlet boundary conditions along the streamwise direction, with the subscripts $\mathrm{D}$ and $\mathrm{B}$ denoting Dirichlet and Neumann boundary constraints, respectively. Finally, $\mathfrak{L}_{a}$ and $\mathfrak{L}_{c}$ impose the electrochemical boundary conditions at the anode and cathode surfaces. The individual contributions to the total loss function are given by
\begin{subequations}
\label{eq:loss_terms}
\begin{align}
\mathfrak{L}_\Omega &=
\frac{1}{N_\Omega} \sum_{k=1}^{N_\Omega}
\Big[
\partial_t \widetilde{C}_A
- D_A \left( \partial_{xx} \widetilde{C}_A + \partial_{yy} \widetilde{C}_A \right)
- \partial_y \widetilde{C}_A \mathbf{v}_{m,A}
+  \partial_x \widetilde{C}_A \mathbf{v}_x
\Big]^2,
\\
\mathfrak{L}_0 &=
\frac{1}{N_0} \sum_{k=1}^{N_0}
\left[
\widetilde{C}_{A,0} - C_{A,0}
\right]^2,
\\
\mathfrak{L}_{\mathrm{D}} &=
\frac{1}{N_\mathrm{D}} \sum_{k=1}^{N_\mathrm{D}}
\left[
\widetilde{C}_{A,\mathrm{D}} - C_{A,0}
\right]^2,
\\
\mathfrak{L}_{\mathrm{B}} &=
\frac{1}{N_\mathrm{B}} \sum_{k=1}^{N_\mathrm{B}}
\left[
\partial_x \widetilde{C}_{A,\mathrm{B}}
\right]^2,
\\
\mathfrak{L}_{a} &=
\frac{1}{N_a} \sum_{k=1}^{N_a}
\Bigg[
- D_A \partial_y \widetilde{C}_{A,\mathrm{s}}
- \widetilde{C}_{A,\mathrm{s}} \mathbf{v}_{m,A}
- \frac{j_{a,0}}{nF}
\left(
\frac{\widetilde{C}_{A,\mathrm{s}}}{C_{A,0}}
\right)^{\gamma_a}\!
\exp\left(
\frac{\alpha_a F \eta_a}{RT}
\right)
\Bigg]^2, \label{eq:lossa}
\\
\mathfrak{L}_{c} &=
\frac{1}{N_c} \sum_{k=1}^{N_c}
\left[
- D_A \partial_y \widetilde{C}_{A,\mathrm{s}}
- \widetilde{C}_{A,\mathrm{s}} \mathbf{v}_{m,A}
\right]^2.
\end{align}
\end{subequations}

The tilde symbol ($\sim$) denotes the neural-network prediction.
At each optimization step, the individual loss components are evaluated on independently drawn mini-batches of randomly sampled collocation points. The sample counts $\{N_\Omega, N_0, N_\mathrm{D}, N_\mathrm{B}, N_a, N_c\}$ denote the number of collocation points used to approximate each contribution to the composite loss at a given iteration. These sample counts do not correspond to a tensor-product discretization of the independent variables (e.g., $N_x \times N_y \times N_t$), but instead represent Monte Carlo sample sizes drawn from the joint input space.

For a generic loss contribution $\mathfrak{L}_\star$, with $\star \in \{\Omega, 0, \mathrm{D}, \mathrm{B}, a, c\}$, a mini-batch
\begin{equation}
\left\{\mathbf{x}^{(k)}_\star\right\}_{k=1}^{N_\star}
=
\left\{\left(x^{(k)}, y^{(k)}, t^{(k)}, \boldsymbol{\theta}^{(k)}\right)\right\}_{k=1}^{N_\star}
\end{equation}
is sampled independently from uniform distributions over the spatio--temporal domain and the prescribed ranges of the conditioning parameters $\boldsymbol{\theta}$. The discrete loss terms defined in Eqs.~\eqref{eq:loss_terms} can therefore be interpreted as Monte Carlo estimates of the corresponding mean-squared residuals, evaluated using standard mini-batch optimization strategies for PINNs \cite{nabian2021efficient,wu2023comprehensive}.

Each mini-batch is assembled into a concatenated input tensor of dimension $N_\star \times 8$, where each row corresponds to an independent realization of a \emph{normalized} input vector $\mathbf{x} \in \mathbb{R}^{8}$.
Here, normalization refers to a deterministic rescaling of all input variables to comparable magnitudes in order to improve numerical conditioning prior to network evaluation.
Specifically, the spatial and temporal coordinates are normalized as $x/L$, $y/d$, and $t/t_{\max}$, while each conditioning parameter $\theta \in \Theta$ is affinely mapped to the unit interval according to
\begin{equation}
\theta \rightarrow 
\frac{\theta - \theta_{\min}}{\theta_{\max} - \theta_{\min}}.
\end{equation}
The neural network outputs a scalar, dimensionless quantity $\mathfrak{n}\!\left(\mathbf{x}\right)$, which is constrained to be non-negative and subsequently rescaled to physical units as
\begin{equation}
\widetilde{C}_A\!\left(\mathbf{x}\right) = \mathfrak{n}\!\left(\mathbf{x}\right) \cdot C_{A,0}.
\end{equation}

This formulation preserves dimensional consistency and ensures that the predicted concentration field is consistently referenced to the inlet concentration scale across different parametric realizations. Negative concentration values are excluded to guarantee a real-valued, unique, and differentiable evaluation of the fractional-order reaction term in the Butler--Volmer-type electrochemical boundary condition, where reaction orders $0<\gamma_a<1$ are employed (see Eq.~\ref{eq:lossa}).

For each approximate solution produced by the PINN, the required partial derivatives in Eqs.~\eqref{eq:loss_terms} are computed using automatic differentiation \cite{paszke2017automatic, baydin2018automatic}. In practice, the independent variables $(x,y,t)$ are treated as differentiable inputs, and derivatives of $\widetilde{C}_A$ with respect to these variables are obtained by back-propagating through the computational graph via the chain rule \cite{cai2021physics}. This avoids truncation and round-off errors associated with numerical differentiation while providing exact derivatives of the network representation.

Training proceeds by minimizing the weighted composite loss in Eq.~\eqref{eq:objective_function} using the Adam optimizer~\cite{kingma2014adam}, with gradients computed via backpropagation and parameter updates performed over mini-batches of collocation points. An adaptive learning-rate schedule is incorporated to enhance robustness across training stages: the optimizer step size $\mathfrak{h}$ is automatically reduced whenever the monitored training loss fails to decrease for $\xi_\mathfrak{h}$ consecutive epochs, according to a multiplicative decay factor $\varsigma$, until a lower bound $\mathfrak{h}_{\min}$ is reached. This strategy enables rapid initial descent while promoting refined convergence during later optimization phases~\cite{bottou2018optimization}. Additionally, gradient-norm clipping is applied to limit excessively large parameter updates and stabilize training in the presence of stiff or highly anisotropic loss landscapes~\cite{pascanu2013difficulty, wang2021understanding}.

\begin{figure}[h!]
    \centering
    \includegraphics[width=0.8\linewidth]{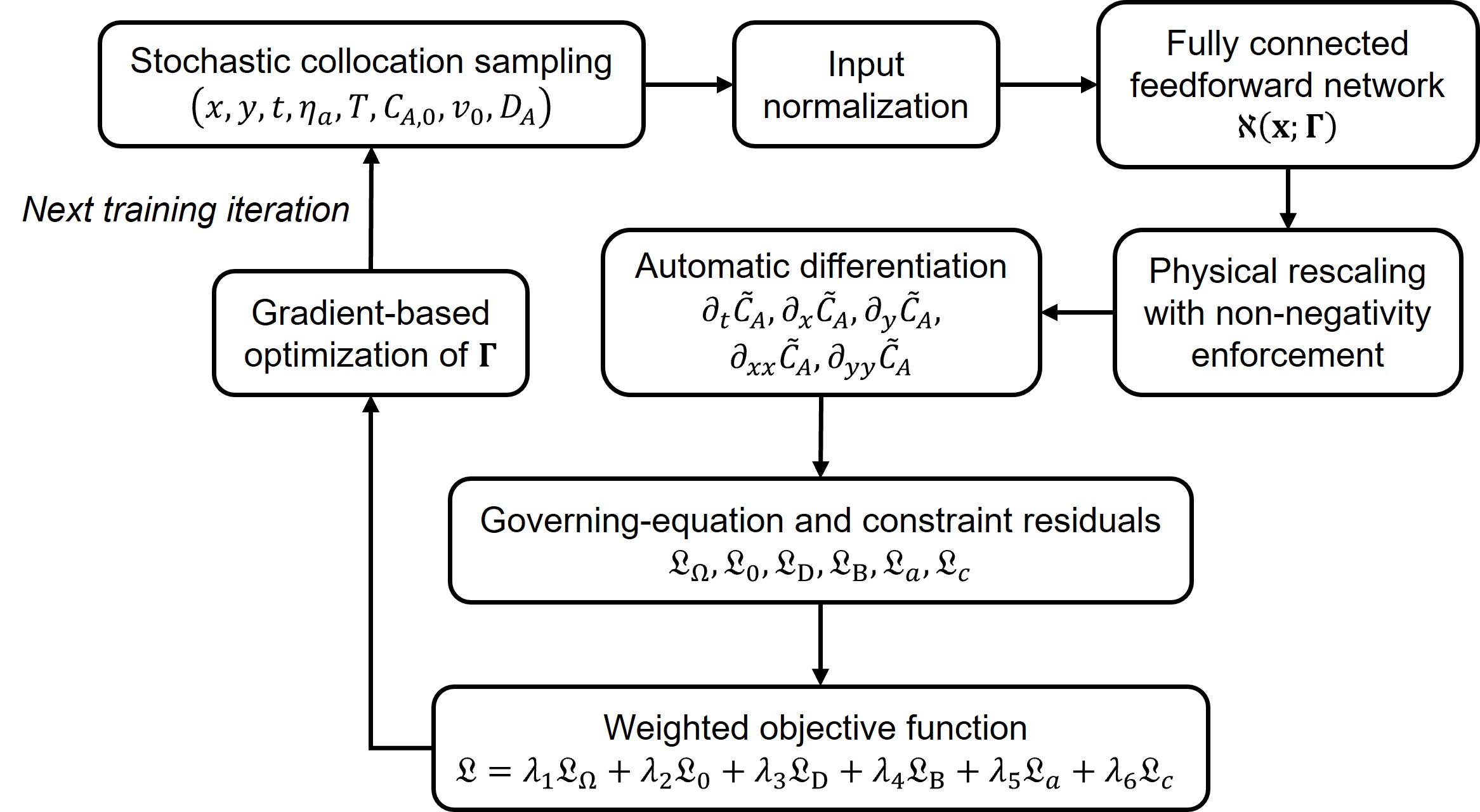}
    \caption{Schematic representation of the PINN training workflow, including stochastic collocation, automatic differentiation, and gradient-based optimization.}
    \label{fig:PINN Diagram}
\end{figure}

Figure~\ref{fig:PINN Diagram} illustrates the overall training pipeline; in practice, each constraint contribution $\mathfrak{L}_\star$ is evaluated on its own independently sampled mini-batch of collocation points, so that interior and boundary sets are not required to share locations or cardinalities. During residual evaluation, only the independent variables $\left(x,y,t\right)$ are treated as differentiable inputs, whereas the conditioning parameters $\boldsymbol{\theta}$ enter as fixed descriptors of each realization, enabling exact derivatives of $\widetilde{C}_A$ with respect to space and time without numerical differencing. The resulting weighted objective is minimized iteratively by repeating this stochastic collocation--evaluation--update cycle for $\Xi$ training epochs, with convergence monitored through the loss evolution. For visualization purposes, the loss curve is smoothed using an exponentially weighted moving average (EWMA),
\begin{equation}
\widetilde{\mathfrak{L}}^{\left(\xi\right)} = \kappa \mathfrak{L}^{\left(\xi\right)} + \left(1-\kappa \right)\widetilde{\mathfrak{L}}^{\left(\xi-1\right)},
\end{equation}
where $\xi$ denotes the training epoch index and $\kappa$ is a smoothing coefficient. This filtering suppresses high-frequency stochastic fluctuations while preserving the overall convergence trend \cite{morales2024exponential}. The EWMA-smoothed loss $\widetilde{\mathfrak{L}}^{\left(\xi\right)}$ is used solely for visualization; optimization is performed using the raw loss $\mathfrak{L}^{\left(\xi\right)}$.

\subsection{Fully connected feedforward neural network}

\begin{figure}[h!]
    \centering
    \includegraphics[width=0.85\linewidth]{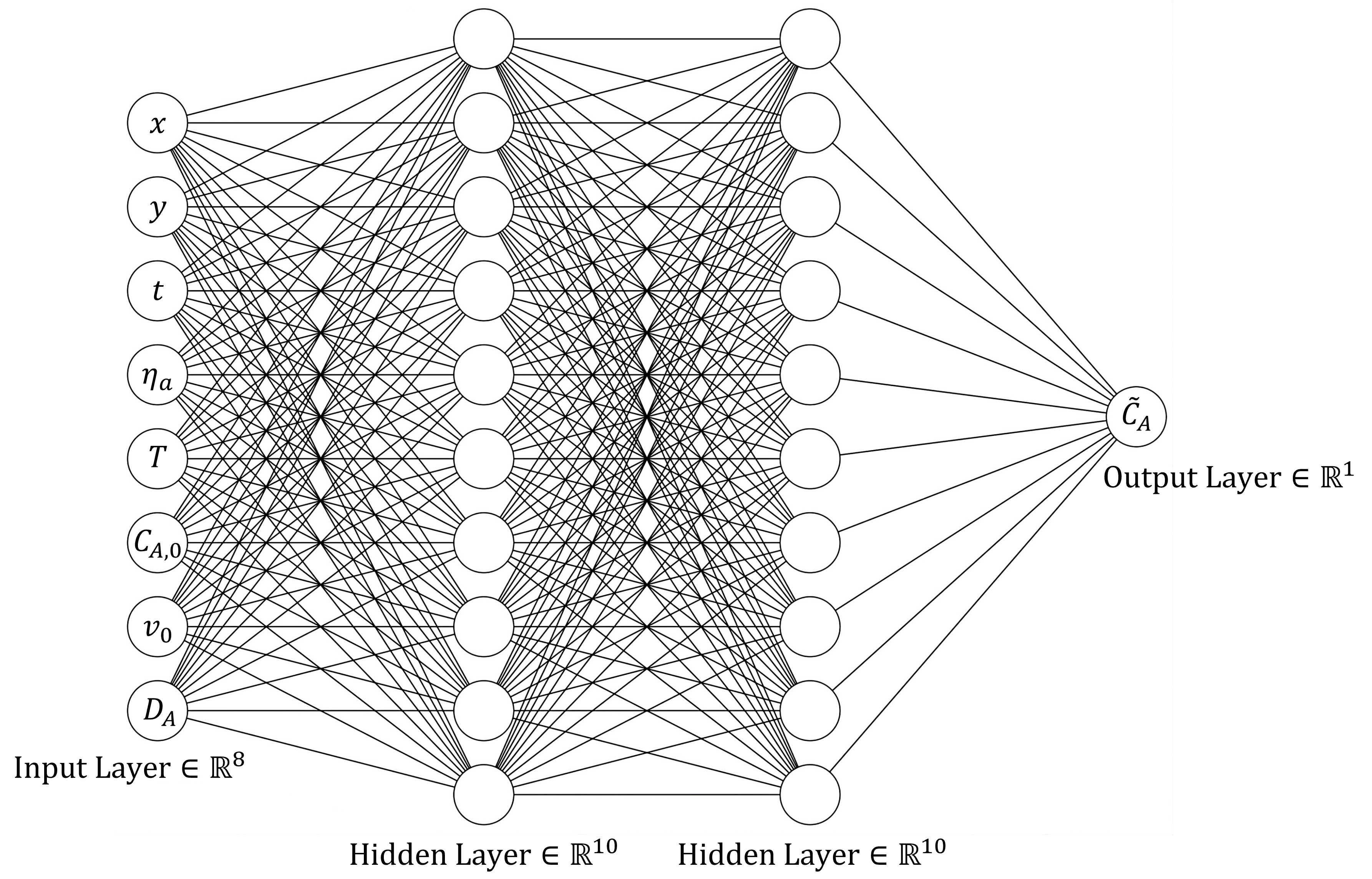}
    \caption{Representative architecture of the fully connected feedforward neural network used in the present PINN, mapping the eight-dimensional input space to the predicted concentration $\widetilde{C}_A$. The diagram is intended for illustration purposes; the specific network depth and number of neurons used in the implementation are described in the PINN design subsection.}
    \label{fig:PINN Diagram 2}
\end{figure}

At the core of the proposed PINN framework lies a fully connected feedforward neural network (FFNN), employed as a universal nonlinear function approximator for the concentration field. As illustrated in Figure~\ref{fig:PINN Diagram 2}, the network consists of an input layer receiving the spatio--temporal coordinates and conditioning parameters, followed by $N_\ell$ densely connected hidden layers and a single-output neuron producing the predicted concentration.

Information propagates strictly in the forward direction through successive affine transformations and pointwise nonlinearities. Denoting by $\mathbf{y}^{\left(0\right)} \equiv \mathbf{x}$ the normalized input vector, the forward mapping through the hidden layers is given by
\begin{equation}
\mathbf{y}^{\left(\ell\right)}
=
\varphi\!\left(
\mathbf{W}^{\left(\ell\right)}\,\mathbf{y}^{\left(\ell-1\right)}+\mathbf{b}^{\left(\ell\right)}
\right),
\qquad
\ell = 1,\ldots,N_\ell,
\end{equation}
where $\mathbf{W}^{\left(\ell\right)}$ and $\mathbf{b}^{\left(\ell\right)}$ denote the weight matrix and bias vector associated with the $\ell$-th layer, respectively, and $\varphi\left(\cdot\right)$ is a smooth nonlinear activation function.

The network output is finally obtained through a linear projection,
\begin{equation}
\mathfrak{n}\!\left(\mathbf{x}\right)
=
\mathbf{W}^{\left(N_\ell+1\right)}\,\mathbf{y}^{\left(N_\ell\right)}
+
\mathbf{b}^{\left(N_\ell+1\right)}.
\end{equation}

In accordance with the classical multilayer perceptron paradigm, the trainable parameters are optimized via gradient-based learning, while in the present PINN formulation the FFNN is embedded within a constrained optimization problem in which the governing transport equation, initial condition, and boundary conditions are enforced through the composite loss function. This coupling enables the network to approximate physically admissible solutions over the entire spatio--temporal domain, while retaining the expressive power and generalization capabilities characteristic of fully connected feedforward architectures \cite{haykin2009neural}.

\subsection{PINN design} \label{sec:pinn_design}
The accuracy and computational performance of the proposed fully connected feedforward PINN are assessed through direct comparison with a conventional explicit finite-difference (FD) model. The FD solver uses second-order central differences for the diffusive terms, first-order upwind discretization for the convective and migrative transport terms, and a forward Euler scheme for time integration. Despite the widespread adoption of PINNs, there is currently no universal or theoretically optimal procedure for selecting their architectural structure or training hyperparameters. Consequently, model design typically follows a pragmatic principle of minimal complexity, whereby the network capacity is chosen to be as small as possible while still achieving the desired predictive accuracy. This strategy is known to promote faster convergence, improved generalization, and reduced susceptibility to overfitting, consistent with the implicit simplicity bias observed in overparameterized neural networks trained with gradient-based optimization \cite{mingard2025deep}.

In this work, the final network architecture, including the number of hidden layers, neurons per layer, and the relative weights assigned to the different loss components, was selected empirically through an extensive series of numerical experiments. These experiments evaluated the minimization behavior and convergence rates of the individual residual terms over a fixed number of training epochs, together with the quantitative agreement between the PINN predictions and the FD reference solutions across the prescribed range of operating parameters. This procedure is consistent with prior analyses showing that imbalanced convergence among loss components can significantly affect training stability and solution accuracy in PINNs \cite{wang2022and}.

Although hyperbolic tangent (tanh) activation functions are commonly employed in classical PINN formulations, the SiLU activation function was adopted in this work based on empirical observations during preliminary training experiments. Compared to tanh, SiLU consistently led to improved training stability and a reduction of spurious oscillations in the predicted solutions \cite{elfwing2018sigmoid}. These effects were particularly noticeable in regimes characterized by strong nonlinearities, such as those arising from Butler–Volmer electrochemical kinetics.

In the final implementation, the network architecture was fixed to a moderately deep fully connected structure consisting of $N_\ell = 4$ hidden layers, each containing 100 neurons. The input layer is eight-dimensional, corresponding to the spatio--temporal coordinates and conditioning parameters, while the output layer is scalar-valued, representing the predicted concentration field $\widetilde{C}_A$.

This configuration provides a total of 400 hidden neurons and was found to offer a suitable balance between expressive capacity and computational efficiency. Smaller networks reduced the training cost but showed insufficient expressive capacity, leading to less consistent concentration magnitudes and poorer enforcement of boundary-driven features. Conversely, increasing the network depth or width beyond this configuration did not yield significant improvements in prediction accuracy, while noticeably increasing training time and sensitivity to optimization.

\section{Results}
\begin{figure}[h!]
    \centering
    \includegraphics[width=1\linewidth]{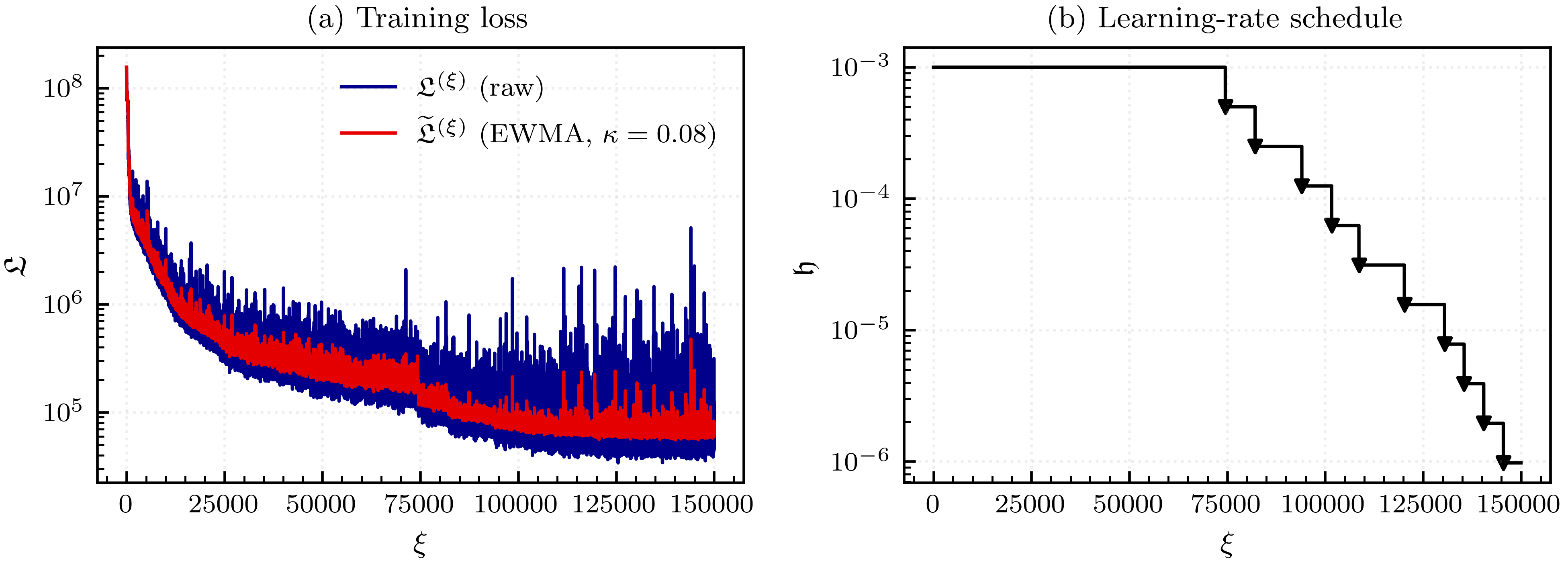}
    \caption{Training diagnostics of the proposed PINN. (a) Evolution of the loss function $\mathfrak{L}^{\left(\xi\right)}$ and its exponentially weighted moving average $\widetilde{\mathfrak{L}}^{\left(\xi\right)}$ versus epoch. (b) Adaptive learning rate $\mathfrak{h}^{\left(\xi\right)}$ as a function of epoch.}
    \label{fig:R1}
\end{figure}

The performance of the PINN is evaluated against the FD reference model under the physical parameters, operating conditions, conditioning ranges, and training settings summarized in Table~\ref{tab:params_pinn}. The training diagnostics shown in Fig.~\ref{fig:R1} indicate stable optimization behavior, characterized by a monotonic decrease of the loss function and a well-regulated adaptive learning-rate schedule. The smoothed loss curve is displayed solely to highlight the underlying convergence trend, whereas parameter updates are performed using the unsmoothed loss values. These observations confirm that the training procedure remains numerically stable and convergent throughout the optimization horizon.

\begin{table}[h!]
\centering
\caption{Physical parameters, operating conditions, conditioning ranges, and training settings used in the FD reference model and the proposed PINN.}
\label{tab:params_pinn}
\begin{tabular}{@{}lll@{}}
\toprule
Symbol & Description & Value / range (units) \\
\midrule

\multicolumn{3}{@{}l}{\textit{Physical constants and electrochemical parameters}} \\
$R$ & Universal gas constant & $8.314~\mathrm{J\,mol^{-1}\,K^{-1}}$ \\
$F$ & Faraday constant & $96485~\mathrm{C\,mol^{-1}}$ \\
$\alpha_a$ & Anodic charge-transfer coefficient & $0.5$ \\
$\gamma_a$ & Anodic reaction order & $0.5$ \\
$j_{a,0}$ & Reference exchange current density (anode) & $10.0~\mathrm{A\,m^{-2}}$ \\
$z_A$ & Charge number of species $A$ & $2$ \\
$n$ & Number of electrons transferred & $1$ \\
$u_A$ & Ionic mobility of species $A$ & $4.04\times 10^{-13}~\mathrm{m^2\,V^{-1}\,s^{-1}}$ \\
\addlinespace

\multicolumn{3}{@{}l}{\textit{Geometry and operating conditions}} \\
$L$ & Channel length & $0.10~\mathrm{m}$ \\
$d$ & Channel height & $1.0\times 10^{-3}~\mathrm{m}$ \\
$\nabla \Phi$ & Approximated potential gradient & $200~\mathrm{V\,m^{-1}}$ \\
$t_{\max}$ & Final simulation time & $20.0~\mathrm{s}$ \\
\addlinespace

\multicolumn{3}{@{}l}{\textit{Conditioning parameter ranges (PINN input space)}} \\
$\eta_a$ & Anodic overpotential & $[0,\,0.3]~\mathrm{V}$ \\
$C_{A,0}$ & Inlet/bulk reference concentration of $A$ & $[100,\,1500]~\mathrm{mol\,m^{-3}}$ \\
$T$ & Temperature & $[273.15,\,500]~\mathrm{K}$ \\
$v_0$ & Characteristic velocity scale ($v_{\max}=v_0/4$) & $[0,\,0.5]~\mathrm{m\,s^{-1}}$ \\
$D_A$ & Diffusivity of species $A$ & $[10^{-9},\,10^{-8}]~\mathrm{m^2\,s^{-1}}$ \\
$\beta$ & Reduced overpotential parameter ($\eta_a/T$) & $[0,\;1.10\times10^{-3}]~\mathrm{V\,K^{-1}}$\\
\addlinespace

\multicolumn{3}{@{}l}{\textit{Training hyperparameters and loss weights}} \\
$N_\star$ & Interior collocation points per epoch & $1000$ \\
$\Xi$ & Training epochs & $150000$ \\
$\lambda_1$ & Weight of $\mathfrak{L}_{\Omega}$ & $10^{3}$ \\
$\lambda_2$ & Weight of $\mathfrak{L}_{0}$ & $10^{2}$ \\
$\lambda_3$ & Weight of $\mathfrak{L}_{\mathrm{D}}$ & $10^{2}$ \\
$\lambda_4$ & Weight of $\mathfrak{L}_{\mathrm{B}}$ & $10^{2}$ \\
$\lambda_5$ & Weight of $\mathfrak{L}_{a}$ & $10^{12}$ \\
$\lambda_6$ & Weight of $\mathfrak{L}_{c}$ & $10^{12}$ \\
$\mathfrak{h}$ & Learning rate (Adam optimizer) & $10^{-3}$ \\
$\xi_\mathfrak{h}$ & Scheduler patience & $5000$ \\
$\varsigma$ & Learning-rate reduction factor & $0.5$ \\
$\mathfrak{h}_{\min}$ & Minimum learning rate & $10^{-8}$\\
\bottomrule
\end{tabular}
\end{table}

Since the Butler--Volmer kinetics depend on temperature and overpotential only through the ratio $\eta_a/T$, we introduce the reduced parameter
\begin{equation}
\beta = \frac{\eta_a}{T},
\end{equation}
which governs the exponential activation term appearing in Eqs.~\eqref{eq:BV_anode}--\eqref{eq:BV_cathode}. This reparameterization captures the combined effect of thermal and electrochemical driving forces through a single scalar variable and avoids redundant exploration of correlated parameters.

\begin{figure}[h!]
    \centering
    \includegraphics[width=1\linewidth]{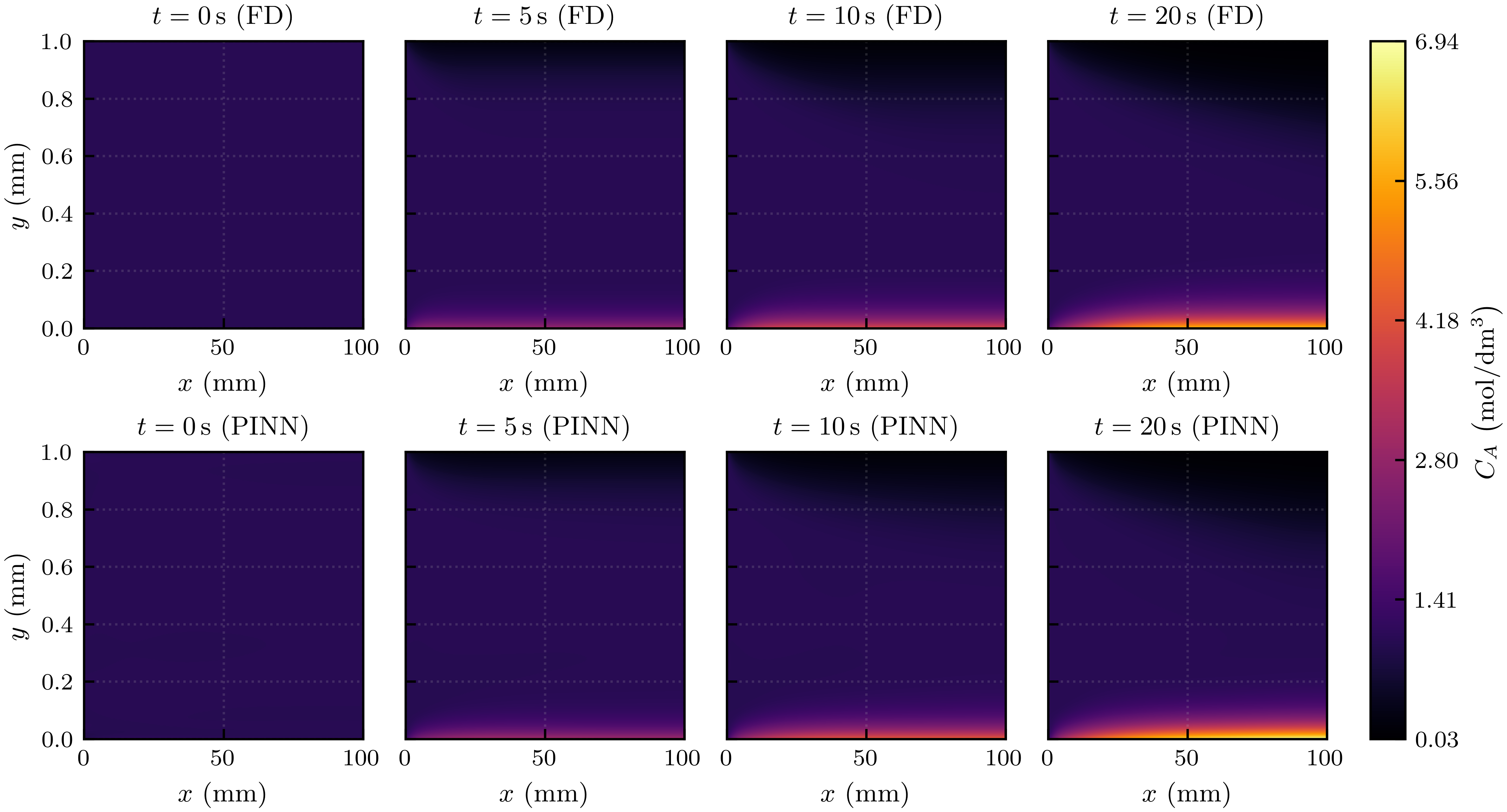}
    \caption{Case~1: Two-dimensional concentration fields of species $A$ predicted by the finite-difference (FD) reference model (top row) and the physics-informed neural network (PINN, bottom row) at the evaluation times $\mathcal{T}=\{0,\,5,\,10,\,20\}\,\mathrm{s}$. Each column corresponds to an instant in time. All FD and PINN maps are rendered using a single global color scale, determined from the combined value range of both models over all times $t\in\mathcal{T}$.}
    \label{fig:C1_maps}
\end{figure}

\begin{figure}[h!]
    \centering
    \includegraphics[width=1\linewidth]{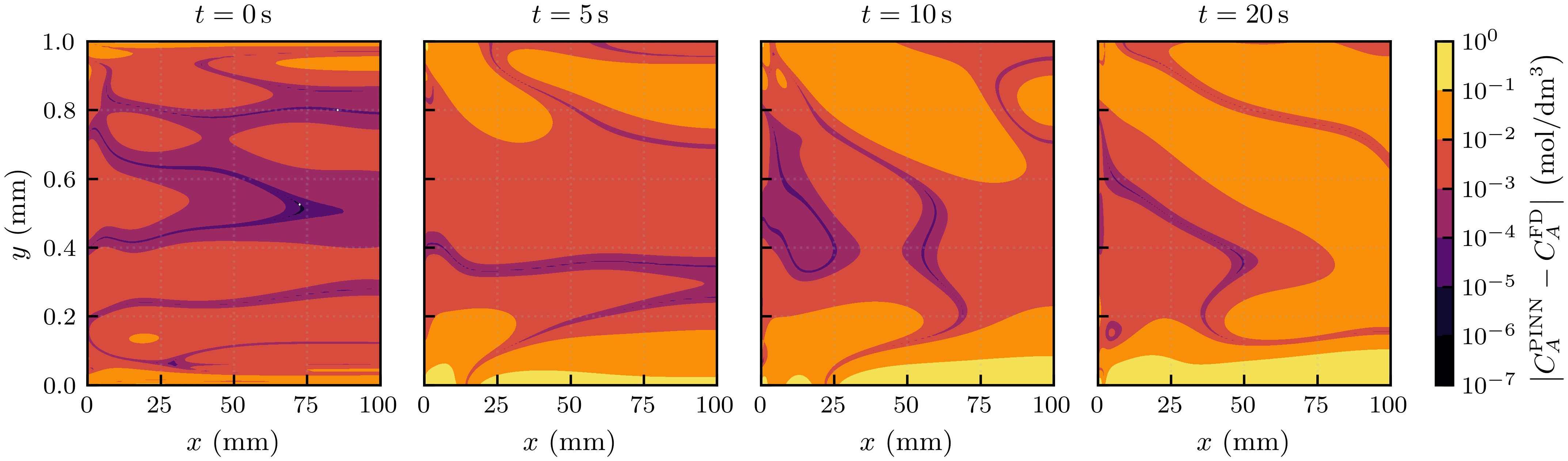}
    \caption{Case~1: Pointwise absolute error maps $\left|C_A^{\mathrm{PINN}}-C_A^{\mathrm{FD}}\right|$ over the domain at the evaluation times $\mathcal{T}=\{0,\,5,\,10,\,20\}\,\mathrm{s}$. A single global logarithmic normalization is used across all panels to resolve discrepancies spanning multiple orders of magnitude.}
    \label{fig:C1_err}
\end{figure}

\begin{figure}[h!]
    \centering
    \includegraphics[width=1\linewidth]{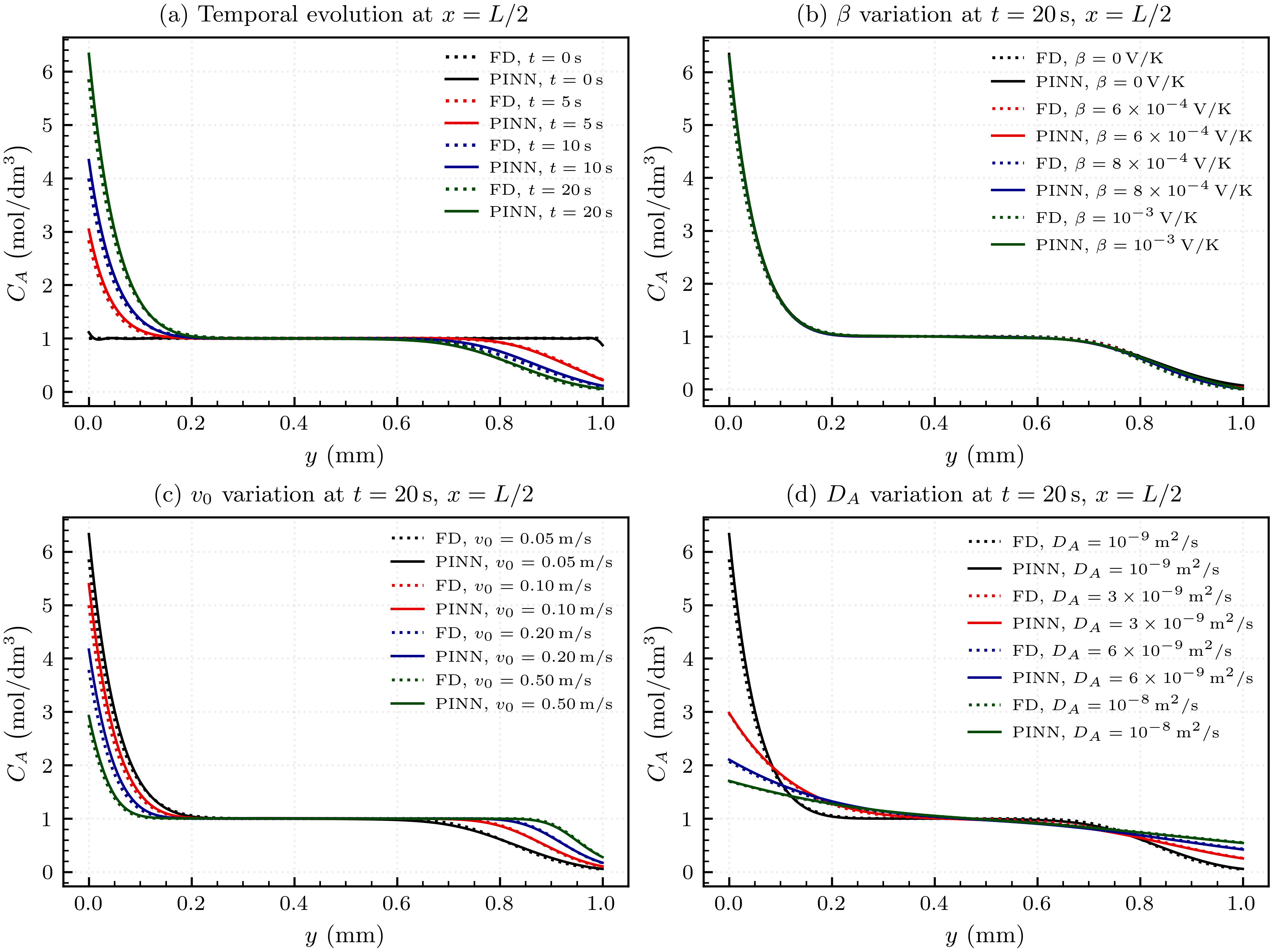}
    \caption{Case~1: Mid-plane concentration profiles at $x=L/2$. (a) Temporal evolution at the reference conditioning vector $\boldsymbol{\theta}$, comparing FD (dotted) and PINN (solid) for all $t\in\mathcal{T}=\{0,\,5,\,10,\,20\}\,\mathrm{s}$. (b--d) Sensitivity of the profiles at the final evaluation time $t=\max(\mathcal{T})$ with respect to the reduced overpotential $\beta=\eta_a/T$, the maximum streamwise velocity $v_0$, and the diffusivity $D_A$, respectively, while keeping the remaining parameters fixed.}
    \label{fig:C1_prof}
\end{figure}

\begin{figure}[h!]
    \centering
    \includegraphics[width=1\linewidth]{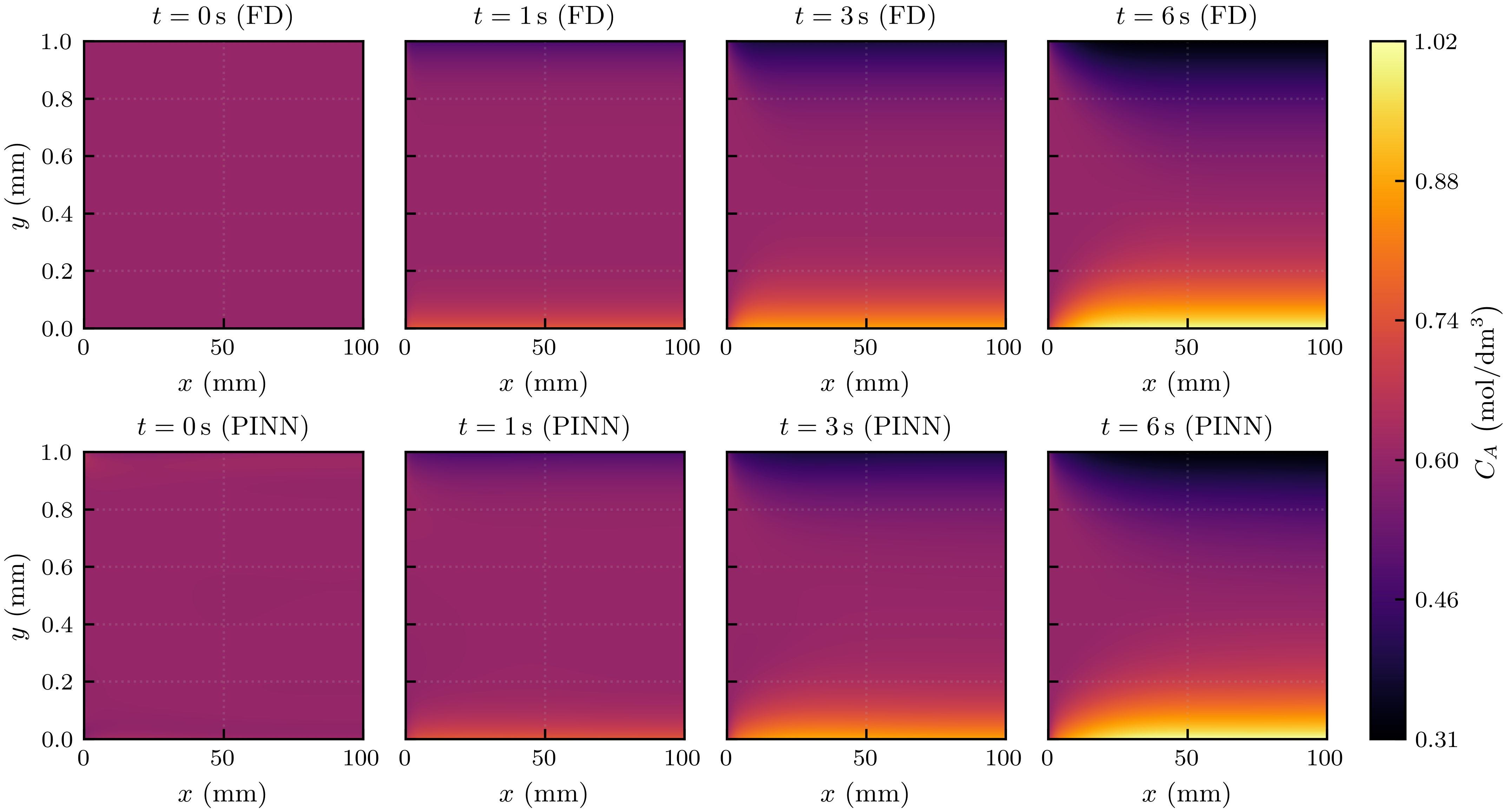}
    \caption{Case~2: Two-dimensional concentration fields of species $A$ predicted by the FD reference model (top row) and the PINN (bottom row) at the evaluation times $\mathcal{T}=\{0,\,1,\,3,\,6\}\,\mathrm{s}$. Each column corresponds to a time instant. All FD and PINN maps are rendered using a single global color scale, determined from the combined value range of both models over all times $t\in\mathcal{T}$.}
    \label{fig:C2_maps}
\end{figure}

\begin{figure}[h!]
    \centering
    \includegraphics[width=1\linewidth]{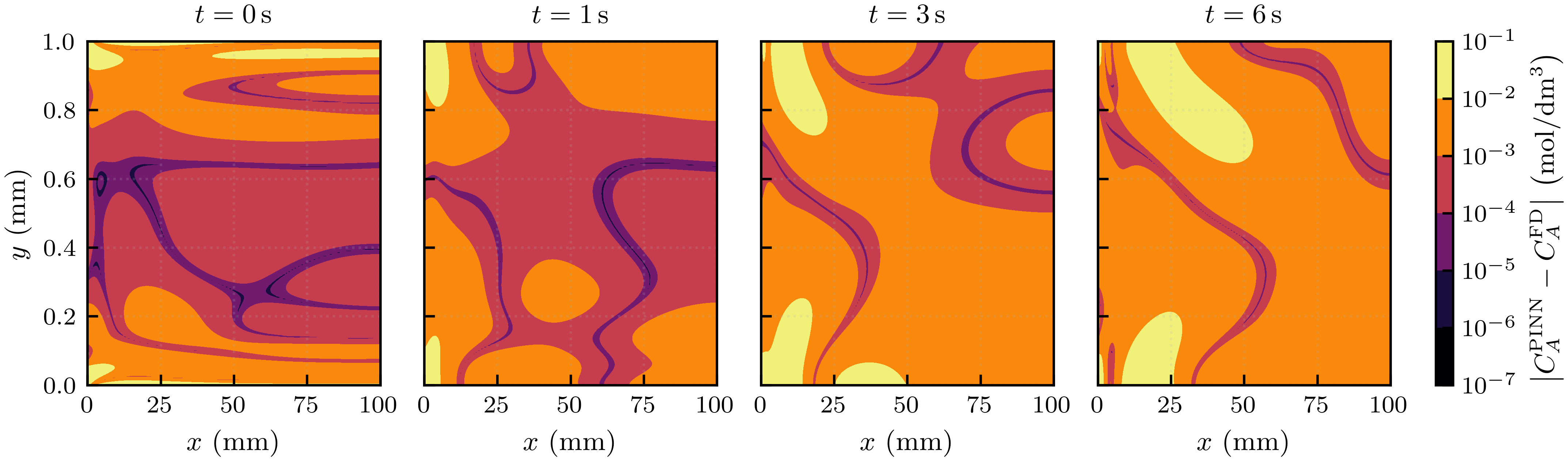}
    \caption{Case~2: Pointwise absolute error maps $\left|C_A^{\mathrm{PINN}}-C_A^{\mathrm{FD}}\right|$ over the domain at the evaluation times $\mathcal{T}=\{0,\,1,\,3,\,6\}\,\mathrm{s}$. A single global logarithmic normalization is used across all panels to resolve discrepancies spanning multiple orders of magnitude.}
    \label{fig:C2_err}
\end{figure}

\begin{figure}[h!]
    \centering
    \includegraphics[width=1\linewidth]{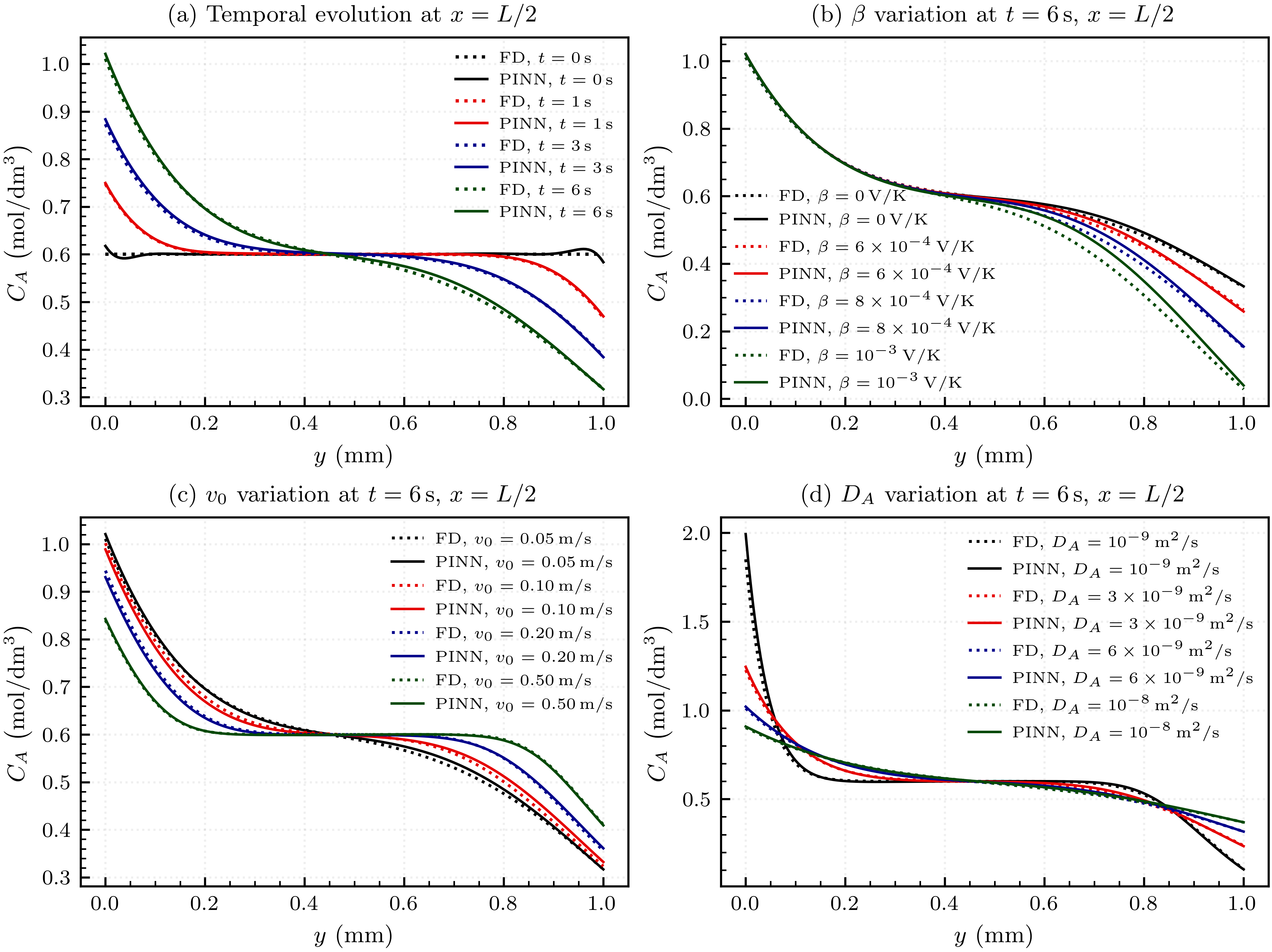}
    \caption{Case~2: Mid-plane concentration profiles at $x=L/2$. (a) Temporal evolution at the reference conditioning vector $\boldsymbol{\theta}$, comparing FD (dotted) and PINN (solid) for all $t\in\mathcal{T}=\{0,\,1,\,3,\,6\}\,\mathrm{s}$. (b--d) Sensitivity of the profiles at the final evaluation time $t=\max(\mathcal{T})$ with respect to $\beta=\eta_a/T$, $v_0$, and $D_A$, respectively, while keeping the remaining parameters fixed.}
    \label{fig:C2_prof}
\end{figure}

Building upon these settings, this section presents a quantitative and qualitative evaluation of the PINN for two representative study cases, each defined by a fixed conditioning vector $\boldsymbol{\theta}$ listed in Table~\ref{tab:study_cases}. For each case, predictions are compared against the FD reference solution in terms of (i) two-dimensional concentration fields (Figs.~\ref{fig:C1_maps} and~\ref{fig:C2_maps}), (ii) transverse concentration profiles evaluated along the mid-plane $x=L/2$ (Figs.~\ref{fig:C1_prof} and~\ref{fig:C2_prof}), and (iii) spatial distributions of the pointwise absolute error (Figs.~\ref{fig:C1_err} and~\ref{fig:C2_err}). For notational clarity, the PINN approximation previously denoted by $\widetilde{C}_A$ is hereafter written as $C_A^{\mathrm{PINN}}$, while the finite-difference reference solution is denoted by $C_A^{\mathrm{FD}}$. The pointwise absolute discrepancy is therefore given by
$$
\left|C_A^{\mathrm{PINN}} - C_A^{\mathrm{FD}}\right|.
$$

We denote by $\mathcal{T}$ the discrete set of evaluation times specified for each case in Table~\ref{tab:study_cases}. These instants are selected to capture both the early transient regime and the subsequent approach toward the coupled migrative--diffusive--convective steady state. In particular, the final evaluation time,
$
t=\max\left(\mathcal{T}\right),
$
is chosen such that the solution has effectively reached convergence in both study cases, ensuring that $\mathcal{T}$ spans the full temporal evolution from initial transients to the asymptotic regime.

To ensure consistent visual interpretation, concentration maps are displayed as shown in Figs.~\ref{fig:C1_maps} and~\ref{fig:C2_maps}, while error distributions (Figs.~\ref{fig:C1_err} and~\ref{fig:C2_err}) are presented using a unified scale to facilitate comparison across time.

\begin{table}[h!]
\centering
\caption{Definition of the representative study cases used for quantitative comparison between the FD reference model and the proposed PINN. Each case corresponds to a fixed conditioning vector $\boldsymbol{\theta} = (\eta_a, T, C_{A,0}, v_0, D_A)$.}
\label{tab:study_cases}
\begin{tabular}{@{}llll@{}}
\toprule
Parameter & Description & Case 1 & Case 2 \\
\midrule

\multicolumn{4}{@{}l}{\textit{Conditioning parameters}} \\

$\eta_a$ & Anodic overpotential (V) 
& $0.10$ 
& $0.10$ \\

$T$ & Temperature (K) 
& $300$ 
& $300$ \\

$C_{A,0}$ & Inlet/bulk reference concentration of $A$ $\left(\mathrm{mol\,m^{-3}}\right)$ 
& $1000$ 
& $600$ \\

$v_0$ & Characteristic velocity scale $\left(\mathrm{m\,s^{-1}}\right)$
& $0.05$ 
& $0.05$ \\

$D_A$ & Diffusivity of species $A$ $\left(\mathrm{m^2\,s^{-1}}\right)$
& $1.0\times10^{-9}$ 
& $6.0\times10^{-9}$ \\

\addlinespace
\multicolumn{4}{@{}l}{\textit{Evaluated time instants}} \\

$\mathcal{T}$ & Evaluation times (s)
& $\{0,\,5,\,10,\,20\}$ 
& $\{0,\,1,\,3,\,6\}$ \\

\bottomrule
\end{tabular}
\end{table}

For each study case, temporal behavior is first examined by fixing $\boldsymbol{\theta}$ and evaluating the solution at all times $t\in\mathcal{T}$ (panel~(a) in Figs.~\ref{fig:C1_prof} and~\ref{fig:C2_prof}). Additional sensitivity analyses are then carried out exclusively for the mid-plane profiles by varying selected components of the conditioning vector ($\beta$, $v_0$, and $D_A$) while keeping the remaining parameters fixed (panels~(b)--(d) in Figs.~\ref{fig:C1_prof} and~\ref{fig:C2_prof}). These parameter variations are performed at a single representative instant, corresponding to the final evaluation time $t = \max\left(\mathcal{T}\right)$, so that comparisons are conducted after the dominant transient dynamics have decayed and the system has reached its converged regime.

Together, these analyses provide a comprehensive characterization of the predictive accuracy of the PINN relative to the FD baseline, demonstrating its ability to reproduce both global transport behavior and localized concentration gradients across the conditioning space.

Beyond the case-based visual comparisons, we quantify the global predictive accuracy of the PINN through a Monte Carlo study over the conditioning space.
Specifically, we draw $M$ independent realizations
$$\left\{\boldsymbol{\theta}^{\left(k\right)}\right\}_{k=1}^{M}.$$
For each sample, we compute a relative discrete space--time error between the PINN and FD solutions evaluated on the same spatial grid
$\mathcal{G}\subset\mathbb{R}^2$ and over the prescribed set of time instants $\mathcal{T}$.

To this end, we introduce the discrete space--time norm
\begin{equation}
\|u\|_{\mathcal{G},\mathcal{T}}^2
:=
\sum_{t\in\mathcal{T}}
\sum_{\mathbf{r}\in\mathcal{G}}
u\left(\mathbf{r},t\right)^2.
\end{equation}
Here $\mathbf{r}$ denotes a generic spatial grid point.

For a given realization $\boldsymbol{\theta}^{(k)}$, the aggregated relative error is defined as
\begin{equation}
\varepsilon^{(k)}=
\frac{
\left\|
C_A^{\mathrm{PINN}}\left(\boldsymbol{\theta}^{\left(k\right)}\right)
-
C_A^{\mathrm{FD}}\left(\boldsymbol{\theta}^{\left(k\right)}\right)
\right\|_{\mathcal{G},\mathcal{T}}
}{
\left\|
C_A^{\mathrm{FD}}\left(\boldsymbol{\theta}^{\left(k\right)}\right)
\right\|_{\mathcal{G},\mathcal{T}}
+
\epsilon
},
\label{eq:mc_rel_error}
\end{equation}
where $\epsilon$ is a small stabilizing constant introduced to prevent division by zero \cite{quarteroni2006numerical}.

Let $\mathcal{E}=\mathbb{E}[\varepsilon]$ denote the expected error.
Its Monte Carlo estimator based on $M$ independent samples reads
\begin{equation}
\hat{\mathcal{E}}_M=\frac{1}{M} \sum_{k=1}^M \varepsilon^{\left(k\right)}
\end{equation}
together with a two-sided $95 \%$ confidence interval obtained from the sample standard error using the Student-$t$ critical value with $M-1$ degrees of freedom.

\begin{figure}[h!]
    \centering
    \includegraphics[width=0.55\linewidth]{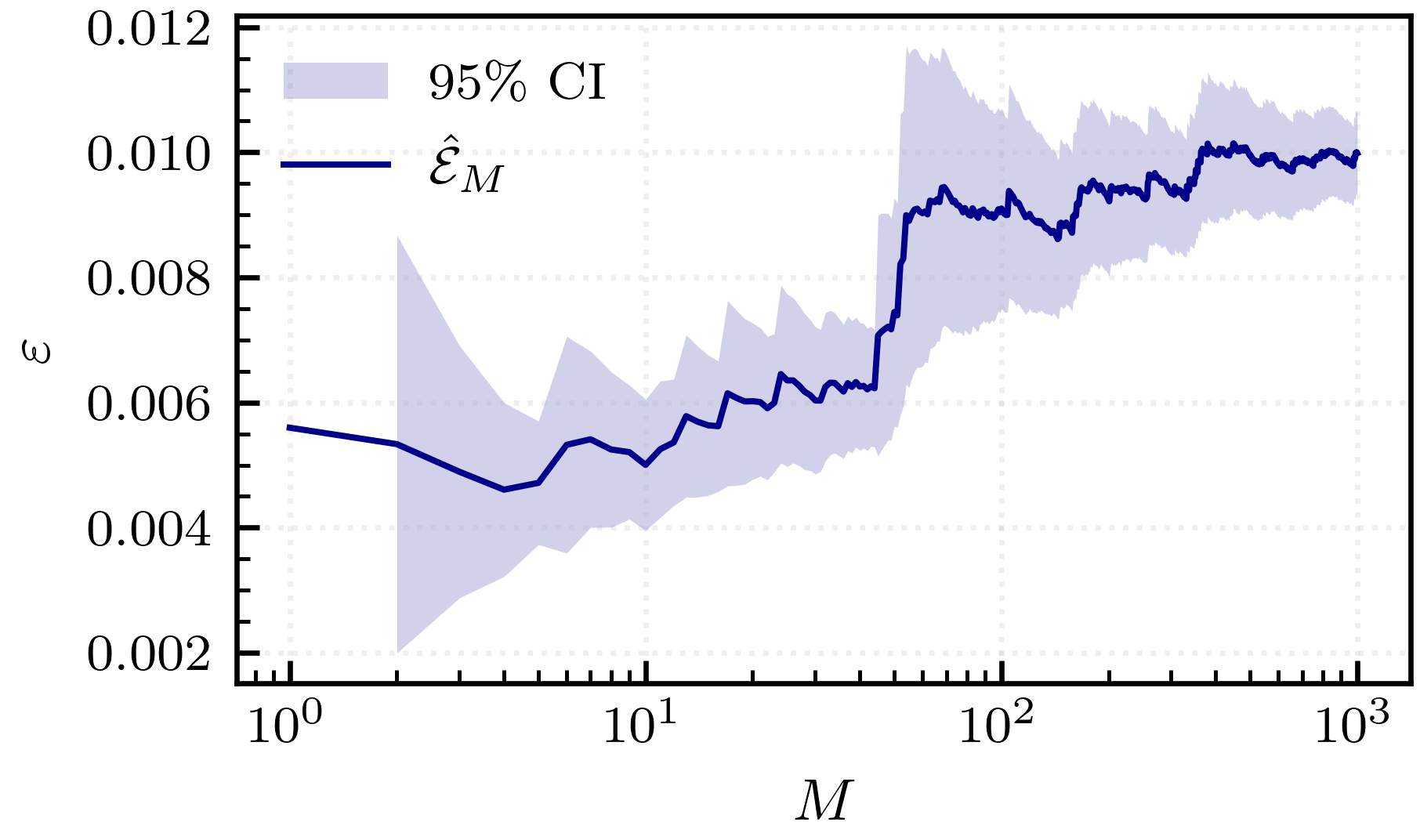}
    \caption{Monte Carlo convergence of the cumulative error estimator $\hat{\mathcal{E}}_{M}$ for the PINN with respect to the FD reference solution over the conditioning space as a function of sample size, $M$. The solid curve shows the running mean of the sample-wise errors $\varepsilon^{(k)}$, while the shaded region denotes the two-sided $95\%$ confidence interval computed from the Student-$t$ distribution. Errors are evaluated over the spatial grid $\mathcal{G}$ $\left(N_x=N_y=201\right)$ and time set $\mathcal{T}=\{0,5,10,15,20\}\,\mathrm{s}$.}
    \label{fig:Error}
\end{figure}

For the evaluation set $\mathcal{T}=\{0,5,10,15,20\}$ and a spatial grid $\mathcal{G}$ of resolution $N_x\times N_y$ with $N_x=N_y=201$ points (corresponding to 200 intervals per direction), the Monte Carlo experiment with $M=1000$ samples yields
$$
\hat{\mathcal{E}}_{M}=(9.99 \pm 0.65)\times 10^{-3},
$$
indicating that the proposed PINN maintains a sub-percent relative error on average over the conditioning domain.
The resulting convergence curve $\hat{\mathcal{E}}_{M}$ and its confidence band (Fig.~\ref{fig:Error}) provide a compact, distribution-level measure of PINN accuracy, complementing the detailed field, profile, and pointwise-error analyses reported for the two representative study cases.

To further probe the robustness of the surrogate beyond average in-domain accuracy, we perform an additional generalization study designed to isolate extrapolation effects and to stress the model near the boundaries of the conditioning space. The same space--time error definition \eqref{eq:mc_rel_error} is retained, and the expected error $\mathcal{E}=\mathbb{E}[\varepsilon]$ is estimated by $\hat{\mathcal{E}}_M$ together with its two-sided $95\%$ confidence interval computed from the Student-$t$ standard error. All evaluations are performed over the same set of discrete time instants $\mathcal{T}$ used in the baseline Monte Carlo experiment. To limit the computational cost of the extended parametric analysis, the spatial grid resolution is moderately reduced while preserving the same number of Monte Carlo samples, as well as the same error definition and evaluation procedure.

First, we consider one-dimensional extrapolation tests in which a single component of the conditioning vector is sampled outside its training range while all remaining components are sampled uniformly within the original domain. Denoting by $\delta$ the extrapolation distance expressed as a fraction of the corresponding training span (reported as a percentage), we generate out-of-domain samples for each parameter in $\boldsymbol{\theta}=(\eta_a, T, C_{A,0}, v_0, D_A)$ at several extrapolation levels $\delta\in[0,\delta_{\max}]$.

For each conditioning parameter $\theta \in \Theta$, extrapolation is applied one-sided above the maximum training value so that sampled values lie in the interval
$$
[\theta_{\max},\,
\theta_{\max} + \delta(\theta_{\max}-\theta_{\min})].
$$
This strategy reflects physically meaningful operating extensions beyond the training regime while avoiding parameter values below the admissible range.

\begin{figure}[h!]
    \centering    \includegraphics[width=1\linewidth]{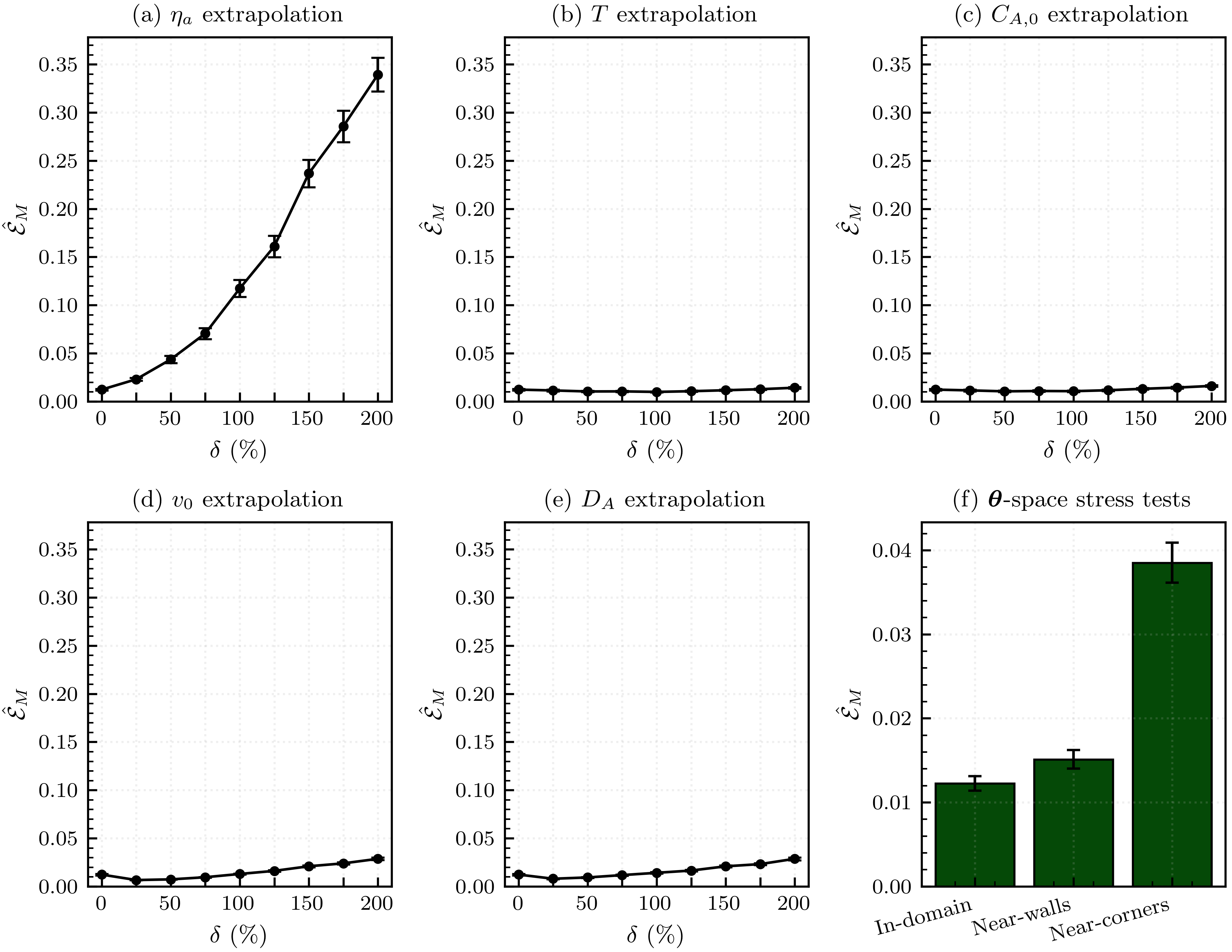}
    \caption{Generalization performance of the proposed PINN under extrapolation and boundary-concentrated sampling in the conditioning space. Panels (a)--(e) show the Monte Carlo estimator $\hat{\mathcal{E}}_M$ as a function of the extrapolation distance $\delta$ for individual conditioning parameters: (a) $\eta_a$, (b) $T$, (c) $C_{A,0}$, (d) $v_0$, and (e) $D_A$. Panel (f) summarizes $\boldsymbol{\theta}$-space stress tests comparing uniform in-domain sampling with boundary-focused sampling strategies (``near-walls'' and ``near-corners''), where parameters are sampled within a $10\%$ boundary band of their respective training ranges. Error bars represent two-sided $95\%$ confidence intervals obtained from $M=1000$ Monte Carlo realizations. Errors are evaluated on the spatial grid $\mathcal{G}$ ($N_x=N_y=101$) and the time set $\mathcal{T}=\{0,5,10,15,20\}\,\mathrm{s}$.}
    \label{fig:ExtrapStress}
\end{figure}

The resulting dependence of $\hat{\mathcal{E}}_M$ on $\delta$ for each parameter is summarized in Fig.~\ref{fig:ExtrapStress}, panels (a)--(e), providing a direct distribution-level measure of how rapidly predictive accuracy degrades as the model is queried progressively farther from the training domain along each coordinate direction \cite{raissi2019physics, krishnapriyan2021characterizing}.

Second, we perform $\boldsymbol{\theta}$-space stress tests that concentrate probability mass near the boundary of the training hyper-rectangle, while still remaining in-domain. In a ``near-walls'' test, one randomly selected parameter is sampled within a thin band adjacent to either its minimum or maximum bound and the remaining parameters are sampled uniformly in-domain, thereby probing sensitivity to single-parameter extremes. In a ``near-corners'' test, all parameters are simultaneously sampled within boundary-adjacent bands, thereby emphasizing rare but challenging combinations in which multiple conditioning variables lie near their limits. The corresponding estimates of $\hat{\mathcal{E}}_M$ are reported in panel (f) of Fig.~\ref{fig:ExtrapStress}. Together, the extrapolation and stress experiments complement the representative case studies and the uniform-domain Monte Carlo benchmark by explicitly testing the stability of the learned surrogate under out-of-distribution queries and boundary-concentrated sampling, which are common in parametric studies and design sweeps \cite{kovachki2023neural}.

\section{Discussion}

Our results demonstrate that the proposed PINN reproduces the electrochemical transport problem in a physically consistent manner across the spatio-temporal and parametric domains considered in this work. Because the model is trained without labeled concentration data and is instead constrained only by the governing equations together with the imposed boundary and initial conditions, the agreement with the finite-difference (FD) reference solution indicates that the network learns a solution that remains faithful to the underlying electrochemical physics. This behavior is particularly relevant for the present system, where migration, diffusion, convection, and nonlinear interfacial kinetics act simultaneously and generate strong concentration gradients near the reactive boundaries.

The concentration maps in Figs.~\ref{fig:C1_maps} and \ref{fig:C2_maps} show that the PINN captures the expected transient redistribution of species $A$ under the combined action of electromigration and anodic consumption. As time evolves, the concentration increases near the cathode and decreases near the anode, which is consistent with the direction of the imposed electric field and with the reactive depletion governed by the Butler--Volmer boundary condition at the anodic surface. The solution progressively approaches a steady state, indicating that the network correctly reproduces both the transient dynamics and the long-time behavior of the system. The presence of concentration boundary layers near the electrodes is also physically consistent with the hydrodynamic configuration of the reactor. In a Poiseuille-driven channel, transverse transport must compete with axial convection and with the interfacial consumption or accumulation of species at the walls, leading to steep concentration gradients near the electrode surfaces. The predicted fields therefore preserve the expected structure of convection–diffusion–migration transport in confined electrochemical geometries.

The comparison between the two representative study cases defined in Table~\ref{tab:study_cases} further confirms the physical consistency of the predictions. In Case~2 the larger diffusivity produces smoother concentration transitions between the electrode surfaces and the bulk electrolyte, reducing the sharpness of the gradients observed near the walls. This behavior reflects the well-known role of diffusion as a gradient-smoothing mechanism and demonstrates that the PINN responds to changes in transport parameters in the same direction as the FD model. Moreover, the close similarity between the FD and PINN concentration maps indicates that both models predict concentration ranges and spatial gradients in good agreement, supporting the conclusion that the PINN reproduces the FD reference solution with sufficient accuracy for the class of electrochemical transport problems considered here. 

The same conclusion is supported by the mid-plane concentration profiles shown in Figs.~\ref{fig:C1_prof} and \ref{fig:C2_prof}. Outside a small discrepancy observed near the $y$-boundaries at the initial time, the FD and PINN curves nearly overlap throughout the transient and near-steady regimes. The localized mismatch close to the boundaries at early times likely reflects the most demanding regions of the conditioning space, where the solution must simultaneously satisfy the initial condition and the nonlinear electrochemical boundary constraint. Since the network is trained over a bounded parameter domain, this behavior suggests that prediction errors may increase slightly near the limits of the conditioning ranges, a tendency that is later confirmed by the $\boldsymbol{\theta}$-space stress tests (Fig. \ref{fig:ExtrapStress}).

The parametric sweeps provide additional insight into the role of the governing parameters. The dependence on the reduced parameter $\beta=\eta_a/T$ highlights the sensitivity of the anodic flux to the exponential activation term in the Butler--Volmer relation. Increasing $\beta$ accelerates the anodic reaction and therefore leads to faster depletion of species $A$ at the anode. The influence of the flow velocity $v_0$ reveals a transport–reaction balance. Larger velocities accelerate the establishment of the quasi-steady concentration profile but also transport electrolyte through the reactor more rapidly, reducing the time available for transverse migration and interfacial consumption. As a result, higher velocities maintain stronger concentration gradients between the electrodes and the bulk while limiting the extent of anodic depletion. A related effect is observed for the diffusion coefficient $D_A$. Increasing diffusivity smooths the concentration gradients across the channel and redistributes species more uniformly throughout the domain. Consequently, lower diffusivity tends to preserve sharper near-electrode concentration gradients, whereas stronger diffusion promotes a more homogeneous concentration field within the reactor.

The error analysis further confirms the quantitative agreement between the PINN and FD solutions. The absolute-error maps in Figs.~\ref{fig:C1_err} and \ref{fig:C2_err} show that discrepancies remain spatially localized and tend to increase during the transient stage, appearing primarily near the final simulation time and in regions close to the cathode where concentration reaches its highest values. Once the solution approaches its asymptotic regime, the error no longer exhibits a sustained increase, suggesting that the largest deviations are associated with the transient evolution rather than with the steady-state concentration pattern. Nevertheless, even in these regions the deviations remain small, and the global Monte Carlo estimate $\hat{\mathcal{E}}_M=(9.99\pm0.65)\times10^{-3}$ indicates that the average relative space--time error remains at the sub-percent level across the conditioning domain. The convergence behavior of the estimator shown in Fig.~\ref{fig:Error} is consistent with the expected stabilization of Monte Carlo averages as the number of samples increases, confirming that the reported error statistics are representative of the entire conditioning domain rather than of a few isolated realizations.


The extrapolation and stress tests summarized in Fig.~\ref{fig:ExtrapStress} provide additional insight into the robustness of the learned surrogate. Among the conditioning parameters, the anodic overpotential $\eta_a$ produces the most pronounced degradation in accuracy during extrapolation, with the mean error increasing by approximately $27.7\times$ at $\delta=200\%$. This behavior is physically expected because $\eta_a$ enters the Butler--Volmer relation through an exponential dependence and therefore induces the strongest nonlinear amplification in the governing equations. By contrast, temperature and inlet concentration remain comparatively benign under extrapolation, exhibiting only modest error increases of about $1.16\times$ and $1.31\times$, respectively, at the same extrapolation level. The flow velocity $v_0$ and diffusivity $D_A$ occupy an intermediate regime, each producing an error increase of approximately $2.34\times$. 

The in-domain stress tests lead to a similar interpretation. Sampling near a single boundary of the conditioning space (``near-walls'') increases the error only moderately relative to uniform in-domain sampling, with a scaling factor of approximately $1.23\times$. In contrast, simultaneous sampling near multiple parameter extremes (``near-corners'') is substantially more challenging, leading to an average error increase of about $3.15\times$. These results indicate that the surrogate remains robust within the training domain while highlighting that predictive stability is most strongly affected by nonlinear amplification associated with the electrochemical driving force.


In terms of computational performance, it is important to distinguish between the offline training cost of the PINN and its subsequent inference efficiency. All computations were performed on a standard personal workstation (AMD Ryzen 5 5600H CPU, NVIDIA GeForce RTX 3050 GPU with 4 GB VRAM), with PINN training carried out in PyTorch using CUDA acceleration. In the present implementation, PINN training required approximately 7806 s ($\approx$ 2.17 h). However, this computational effort is incurred only once during the offline training stage. Once trained, the PINN provides full-field predictions for the concentration $C_A$ over the spatial grid $\mathcal{G}$ $(N_x=N_y=201)$ with an average inference time of $1.57 \pm 0.26$ s over 100 representative operating conditions $\theta \in \Theta$, with runtimes ranging from 1.06 to 2.82 s. These cases comprise the two study cases defined in Table~\ref{tab:study_cases}, together with additional randomly sampled configurations within the training domain. These timings correspond to the reconstruction of the solution at the evaluation times $\mathcal{T}=\{0,5,10,15,20\}\,\mathrm{s}$, which are introduced exclusively for computational benchmarking and differ from the case-specific time sets defined in Table~\ref{tab:study_cases}. In contrast, the FD solver required on average $8.39 \pm 3.57$ s per simulation over the same set of operating conditions, with runtimes ranging from 1.32 to 14.41 s. Thus, the trained PINN achieves an average speedup of approximately $5.34\times$ for this simplified two-dimensional system. This advantage is expected to become increasingly significant for higher-dimensional and more complex multiphysics problems, where repeated conventional simulations would be considerably more expensive.

From an application standpoint, the primary advantage of the proposed PINN framework lies not in reducing the cost of a single simulation, but in providing a continuous parametric representation of the electrochemical reactor once training has been completed. Although the cost of an individual FD simulation remains modest in the present setting, exploring the multidimensional conditioning space would require repeatedly solving the governing equations using conventional numerical methods. In contrast, the trained PINN provides immediate predictions for new operating conditions within the learned domain, enabling rapid parametric sweeps over $\boldsymbol{\theta}$ and facilitating tasks such as sensitivity analysis, optimization, and design exploration. This capability becomes increasingly valuable as model complexity increases, for example when incorporating multicomponent transport, more detailed electrochemical kinetics, or more complex reactor geometries \cite{jarvey2022ion}.

Finally, several limitations of the present formulation should be noted. The current model explicitly resolves only a single electroactive species and assumes fully developed laminar Poiseuille flow together with a spatially uniform electric field. These simplifications are appropriate for demonstrating the feasibility of a data-free PINN formulation for electrochemical transport but do not represent the full complexity of practical electrochemical systems. Future developments may incorporate multicomponent transport, self-consistent potential fields, and time-dependent forcing conditions. Nonetheless, the present results serve to demonstrate that PINNs can accurately reproduce nonlinear electrochemical transport phenomena and provide a promising framework for fast parametric reactor modeling.

\section{Conclusions}

This work presents a physics-informed neural network (PINN) for modeling a transient two-dimensional electrochemical flow reactor with diffusion, migration, convection, and nonlinear anodic Butler--Volmer kinetics. Without using labeled concentration data, the model reproduces the finite-difference reference solutions with strong qualitative and quantitative agreement across representative operating conditions and throughout the training domain. The predicted concentration fields preserve the expected physical behavior of the system, including anodic depletion, cathodic accumulation driven by electromigration, diffusive smoothing of concentration gradients, convective transport under Poiseuille flow, and the progressive approach toward steady state.

Comparison with the finite-difference model shows that the proposed PINN is not only visually consistent with the reference solution but also statistically accurate at the distribution level, yielding a sub-percent average relative error across the conditioning domain. The additional extrapolation and $\boldsymbol{\theta}$-space stress tests further revealed that generalization is parameter-dependent, with the anodic overpotential representing the most demanding direction due to its exponential role in the Butler--Volmer boundary condition. These results provide a useful characterization of the regions of the conditioning space where the surrogate remains most robust and where additional training data or architectural refinements may be beneficial.

In practical terms, the primary value of the PINN framework lies in its ability to provide a fast parametric surrogate once training has been completed, a task that can always be done offline. For the computational benchmark, this corresponds to an average speedup of approximately $5.34\times$, or an $81.3\%$ reduction in runtime relative to the FD solver. While the computational cost of a single finite-difference simulation remains modest for the simplified system considered here, exploring the multidimensional conditioning space using conventional solvers would require repeatedly resolving the governing equations. In contrast, the trained PINN provides immediate predictions for new operating conditions within the learned domain. In this sense, the model can be interpreted as a first step toward a digital twin of the electrochemical reactor: a physics-constrained computational representation capable of rapidly predicting concentration fields under varying operating conditions. This may in turn support tasks such as parametric screening, optimization, and design exploration entirely in silico.

The present formulation should therefore be viewed as a foundational demonstration that PINNs can be meaningfully applied in electrochemical modeling, even in the absence of training data and in the presence of nonlinear electrochemical boundary conditions. Despite the simplifying assumptions of the present model, the results demonstrate that physics-informed machine-learning-based surrogates can become a useful new tool for electrochemical reactor modeling, particularly when repeated parametric exploration of operating conditions is required.

%
%

\section*{Acknowledgments}
MM acknowledges the generous support of Tecnológico de Monterrey received under the Faculty of Excellence program.

\bibliographystyle{unsrt}  

\bibliography{references}

\end{document}